\documentclass[a4,11pt]{article}
\usepackage{amsmath,amsthm,amssymb,braket,bm}
\usepackage{cite}

\usepackage{jcappub}

\date{\today}

\title{Point splitting renormalization of Schwinger induced current in de Sitter spacetime}
\author[a, b]{Takahiro Hayashinaka}
\author[a, b, c]{and Jun'ichi Yokoyama}
\affiliation[a]{Research Center for the Early Universe (RESCEU), Graduate School of Science, 
The University of Tokyo, \\
Bunkyo, Tokyo, 113-0033, Japan}
\affiliation[b]{Department of Physics, Graduate School of Science, 
The University of Tokyo, \\
Bunkyo, Tokyo, 113-0033, Japan}
\affiliation[c]{Kavli Institute for the Physics and Mathematics of the Universe (Kavli IPMU), 
WPI, UTIAS, The University of Tokyo, \\
Kashiwa, Chiba, 277-8583, Japan}

\emailAdd{hayashinaka@resceu.s.u-tokyo.ac.jp}
\emailAdd{yokoyama@resceu.s.u-tokyo.ac.jp}

\abstract{
The covariant and gauge invariant calculation of the current expectation value in the homogeneous electric field 
in 1+3 dimensional de Sitter spacetime is shown. 
The result accords with previous work obtained by using adiabatic subtraction scheme. 
We therefore conclude the counterintuitive behaviors of the current in the infrared (IR) regime such as IR hyperconductivity and negative current 
are not artifacts of the renormalization scheme, but are real IR effects of the spacetime. 
}

\keywords{Schwinger effect, de Sitter spacetime}
\arxivnumber{1603.06172}

\begin{document}
\maketitle

\section{Introduction}
Investigation of quantum field theory in curved spacetime background has a long history. 
We cannot put enough emphasis on the importance of the study in the context of cosmology \cite{birrell1984quantum, mukhanov2007introduction}. 
Inflationary background, which is approximately described by de Sitter spacetime, may be the most interesting and relevant subject 
(for a review of inflationary cosmology, see e.g. \cite{doi:10.1142/S0218271815300256}). 
One of the greatest achievements in the inflationary cosmology is the prediction of the primordial perturbations 
which is supposed to become a seed of all the structure in the later universe. 
Their calculation is entirely based on quantum field theory in curved spacetime and the agreement with observations infers the correctness of the approach. 
Not only the scalar perturbation but also the vector and tensor perturbations can be generated from the quantum fluctuations in the inflationary universe. 
The cosmic microwave background (CMB) observations basically espouse the generation of the primordial scalar perturbation. 
The primordial tensor perturbation has also been investigated and detection of the primordial gravitational waves is awaited. 

The primordial vector perturbation is usually less significant since it only has decaying modes. 
On the other hand, the observations of galactic/extragalactic magnetic field 
\cite{plaga1995detecting,0004-637X-682-1-127,neronov2010evidence,2041-8205-747-1-L14,2041-8205-744-1-L7,tashiro2014search,Akahori01062014,Schnitzeler01112010} 
indicate the existence of large scale magnetic field in extragalactic scale whose origin is yet to be clarified. 
The inflationary magnetogenesis is a serious candidate since it may be able to produce coherent magnetic fields on large scale, 
especially, extragalactic scales ($\sim$Mpc). 
Thus it is motivated to modify the standard theory which is conformally invariant and hence no long-wave perturbations are expected 
so that the primordial vector perturbation can be generated during inflationary era. 
Among many proposed mechanisms of inflationary magnetogenesis, one of the most actively investigated models is the 
so-called $f^2FF$ model \cite{ratra1992cosmological,PhysRevD.37.2743,PhysRevD.69.043507,PhysRevD.70.083508,martin2008generation} 
where the kinetic term has a nontrivial time dependence. 
However, it suffers from the backreaction problem, namely, overproduction of the electric fields 
which occurs if one tries to avoid the strong coupling problem of the theory \cite{demozzi2009magnetic}. 

A natural consequence of the strong electric field is pair production 
of charged particles known as the Schwinger effect \cite{schwinger1951gauge}, 
which is an example of the nonperturbative effect of the quantum field theory. 
Recently, several studies on this subject in de Sitter spacetime have been done 
\cite{PhysRevD.49.6343,1475-7516-2014-04-009, Cai:2014qba, kobayashi2014schwinger, Stahl:2015gaa, Bavarsad:2016cxh}. 
Their motivations vary widely from false vacuum decay and bubble nucleation 
to a thermal interpretation of particle production or cosmological consequences including magnetogenesis. 
The particle production rate can be calculated once the 
Bogoliubov coefficients, which constitute the connection matrix between the in-vacuum and the out-vacuum mode function, are obtained. 
The real obstacle is the lack of the proper definition of the out-vacuum state at an arbitrary time in curved spacetime.  
In order to estimate the back reaction of the Schwinger effect, we do not
need to calculate the particle production rate itself but instead the time evolution of the expectation value of the induced current would suffice.
This is the strategy adopted in \cite{1475-7516-2014-04-009, kobayashi2014schwinger, Stahl:2015gaa, Bavarsad:2016cxh}. 
So far, the vacuum expectation value of the current has been calculated for scalar quantum electrodynamics (QED) in $1+1$ dimensional de Sitter spacetime 
\cite{1475-7516-2014-04-009}, in $1+2$ dimension \cite{Bavarsad:2016cxh}, in $1+3$ dimension \cite{kobayashi2014schwinger} and for spinor QED in $1+1$ dimensional de Sitter spacetime 
\cite{Stahl:2015gaa}, in $1+3$ dimension \cite{Hayashinaka:2016qqn}. 
The adiabatic subtraction scheme up to second order was employed to obtain a regularized current in \cite{kobayashi2014schwinger}, 
while it is necessary to do the fourth order regularization to obtain finite expression of the energy momentum tensor. 
Hence it is desired to compare their result with those using other renormalization schemes.

Another issue is that the previous calculation also used momentum cutoff to control the divergence, which breaks the gauge invariance. 
It is well-known that the gauge symmetry ensures the renormalizability of QED theory. 
Actually, cutoff regularization brought about unrenormalizable divergence(s) to the theory. 
To avoid this problem, we have to regularize the divergence in a gauge-invariant way. 
Dimensional regularization is often used for this purpose but it does not work in our case, as it does not control the divergence. 
Instead, we can make use of the point splitting technique \cite{birrell1984quantum,dewitt1975quantum} .
In this scheme, the covariant point separation is used to control the divergence. 

Our aim in this paper is to perform the point-splitting renormalization of the vacuum expectation value of the scalar current in 
$1+3$ dimensional de Sitter spacetime in a covariant and gauge invariant way. 
We choose the physically same background gauge field seen in \cite{kobayashi2014schwinger}.

The construction of this paper is as follows. 
In Sec.~\ref{Sec:calculation}, we will introduce the method of calculation and perform it. 
In Sec.~\ref{Sec:result}, the properties of the result will be investigated. 
In Sec.~\ref{Sec:discussion}, a possible physical interpretation for the result is given.
Finally, summary and conclusion are given in Sec.~\ref{Sec:conclusion}. 

\section{Point splitting renormalization of Schwinger induced current}\label{Sec:calculation}
\subsection{Set up}
We investigate the scalar QED theory consisting of an U(1) gauge field $A_\mu(\bm x,t)$ and a complex scalar field $\phi(\bm x,t)$ with charge $e$
in de Sitter space 
\begin{equation}
\mathrm ds^2 =g_{\mu\nu}\mathrm dx^{\mu}\mathrm dx^{\nu}= -\mathrm dt^2 + \mathrm e^{2 Ht}\mathrm d\bm x^2 = a^2(\eta) (-\mathrm d\eta^2 + \mathrm d\bm x^2), 
\quad a(\eta) = \frac{1}{1-H\eta},  \label{desitter}
\end{equation}
whose action is given by 
\begin{equation}
S = \int \mathrm d^4x\sqrt{-g} 
\left( -\frac{1}{4}g^{\mu\alpha}g^{\nu\beta}F_{\mu\nu}F_{\alpha\beta} 
- g^{\mu\nu}(D_{\mu}\phi)^\dagger D_\nu\phi -m^2\phi^\dagger\phi \right),
\end{equation}
with $D_\mu = \partial_\mu + ieA_\mu$ and $F_{\mu\nu}=\nabla_\mu A_\nu-\nabla_\nu A_\mu$ . 
We treat the gauge field $A_\mu$ as a background field giving rise to a constant electric field. 
That is, we choose its configuration as 
\begin{equation}
A_\mu(x) = (0,0,0,A_z(\eta)), \quad A_z(\eta) = -\frac{E}{H}(a(\eta)-1). 
\end{equation}
As defined in (\ref{desitter}) we take the scale factor as $a(\eta)  = \mathrm e^{Ht} = (1-H\eta)^{-1}$ 
so that we can obtain the explicit Minkowski ($H\to 0$) limit, $A_z(\eta) \to -Et$. 

The local current operator is defined by 
\begin{equation}
\hat J_\mu(x) \equiv ie \hat\phi^\dagger(x) \overleftrightarrow{D_\mu} \hat\phi(x) 
= ie \{\hat\phi^\dagger(x) D_\mu \hat\phi(x) - (D_\mu\hat\phi(x))^\dagger \hat\phi(x)\}. 
\end{equation}
As the vacuum expectation value of this operator is divergent,
we adopt the point separation $x^\mu \rightarrow x^\mu \pm \epsilon^\mu$ to control the divergence and 
renormalize it in a gauge-invariant manner.

\subsection{gauge-invariant two-point current operator}
The gauge-invariant two-point current operator with symmetric point separation is given by 
\begin{equation}\label{Jxepsilon}
\hat J_\mu(x;\epsilon) \equiv 
ie \,\exp\left[{-ie\int_{x-\epsilon}^{x+\epsilon} \mathrm dx^\mu A_\mu } \right]
\hat\phi^\dagger(x+\epsilon) \overleftrightarrow{D_\mu} \hat\phi(x-\epsilon), 
\end{equation}
which is invariant under the gauge transformation with an arbitrary function $\Gamma(x)$, 
\begin{equation}
\hat\phi(x) \to \mathrm e^{-ie\Gamma(x)}\hat\phi(x), \quad 
\hat\phi(x)^\dagger \to \mathrm e^{+ie\Gamma(x)}\hat\phi^\dagger(x), \quad
A_\mu (x) \to A_\mu + \Gamma_{,\mu}(x). 
\end{equation}
Note that the covariant derivative is transformed as $D_\mu\hat\phi(x) \to \mathrm e^{-ie\Gamma(x)}D_\mu\hat\phi(x)$ and changes the overall phase. 
This is canceled by the prefactor $\exp[{-ie\int_{x-\epsilon}^{x+\epsilon} \mathrm dx^\mu A_\mu }]$ 
which will be unity when the coincidence limit $\epsilon \to 0$ is taken. 
Of course, we can recover the locality of the current operator in the coincidence limit, 
\begin{equation}
\hat J(x) = \lim_{\epsilon\to 0} \hat J(x;\epsilon)
\end{equation}
We can also separate the vacuum expectation value $\displaystyle \lim_{\epsilon\to 0} \braket{\hat J(x;\epsilon)}$ into 
the $\epsilon$-dependent divergent terms and the $\epsilon$-independent finite terms as we will see below. 
This fact indicates that the divergence has an ultraviolet (UV) nature and can be absorbed by renormalization of the charge $e$ and the field redefinition. 

The mode decomposition of the quantized scalar field is given by 
\begin{equation}
\hat \phi(x) = \frac{1}{a}\int\frac{\mathrm d^3k}{(2\pi)^3}\mathrm e^{i\bm k\cdot \bm x} 
\left( \chi_{\bm k}(\eta) \hat a_{\bm k} + \chi_{\bm k}^\ast(\eta)\hat b_{-\bm k}^\dagger \right) ,
\end{equation}
where $\chi_{\bm k}(\eta)$ is a canonical mode function. 
It satisfies the following field equation 
\begin{equation}
\left[\frac{\partial^2}{\partial\eta^2} 
+ \left(m^2+\left(\frac{eE}{H}\right)^2-2H^2\right)a^2 - 2\frac{eE}{H}\left(k_z+\frac{eE}{H}\right)a + k^2+2\frac{eE}{H}k_z+\left(\frac{eE}{H}\right)^2 
 \right]\chi_{\bm k}(\eta)= 0, 
\end{equation}
which is solved in terms of the Whittaker function $W_{\kappa,\mu}(z)$ \cite{gradshteyn2007} as 
\begin{equation}
\chi_{\bm k}(\eta) = \frac{\mathrm e^{i\pi\kappa/2}}{\sqrt{2p}}  W_{\kappa,\mu}(z), 
\end{equation}
with 
\begin{equation}
z \equiv -2i\frac{p}{aH},\; \kappa \equiv -iL\frac{p_z}{p},\; \mu \equiv \sqrt{\frac{9}{4}-L^2-M^2},\; 
L \equiv \frac{eE}{H^2},\; M \equiv \frac{m}{H}. 
\end{equation}
Here we have introduced shifted momentum $\bm p = (k_z,k_y,k_z+HL)$. 
The creation and annihilation operators $a_{\bm k}, \, b_{\bm k},\,a_{\bm k}^\dagger, \, b_{\bm k}^\dagger$ 
satisfy the canonical commutation relations 
$[a_{\bm k},a_{\bm k^\prime}^\dagger] = [b_{\bm k},b_{\bm k^\prime}^\dagger] = (2\pi)^3\delta^{(3)}(\bm k-\bm k^\prime)$ and 
$(\mathrm{others})=0$. 

Choosing a straight line as the integration contour in (\ref{Jxepsilon}), we obtain
\begin{equation}
\braket{\hat J_z(x;\epsilon)} = 
-2e\left(\frac{a_+}{a_-}\right)^{-iL\frac{\Delta z}{\Delta \eta}}\int\frac{\mathrm d^3p}{(2\pi)^3}\frac{\mathrm e^{-i\bm p\cdot \Delta x}}{a_+a_-}
(p_z-\bar a HL)\chi_{\bm k}(\eta+\frac{\Delta\eta}{2}) \chi_{\bm k}^\ast(\eta-\frac{\Delta\eta}{2}), 
\end{equation}
where we have used $\epsilon^\mu = (\Delta\eta/2,\Delta\bm x/2)$ and introduced $a_\pm = [1-H(\eta\pm\Delta\eta/2)]^{-1}$ and $\bar a = (a_+ + a_-)/2$. 
We can make use of the Mellin-Barnes representation for the Whittaker function \cite{gradshteyn2007} to evaluate this quantity, 
\begin{equation}\label{MBrep}
W_{\kappa,\mu}(z) =\int_{C_s}\frac{\mathrm ds}{2\pi i} z^s \mathrm e^{-z/2} \frac{\Gamma(s-\kappa)\Gamma(-s-\mu+\frac{1}{2})\Gamma(-s+\mu+\frac{1}{2})}{\Gamma(\frac{1}{2}-\kappa-\mu)\Gamma(\frac{1}{2}-\kappa+\mu)}, 
\end{equation}
where the integration contour $C_s$ runs from $-i\infty$ to $i\infty$ and is taken to separate the poles of $\Gamma(s-\kappa)$ ($s=\kappa-n$, $n=0,1,2,\cdots$) from 
those of $\Gamma(-s-\kappa-\mu+\frac{1}{2})\Gamma(-s-\kappa+\mu+\frac{1}{2})$. 
After substituting (\ref{MBrep}), the expectation value reads 
\begin{equation}\begin{split}
&\braket{\hat J_z(x;\epsilon)} = 
-\frac{e}{(2\pi)^3a_+a_-}\left(\frac{a_+}{a_-}\right)^{-iL\frac{\Delta z}{\Delta \eta}}
\int_0^\infty \mathrm dp \int_{-1}^{1} \mathrm d\xi \int_0^{2\pi}\mathrm d\varphi \,\mathrm e^{-ip\eta-i\bm p\cdot \Delta x} 
 \int_{C_s}\frac{\mathrm ds}{2\pi i} \int_{C_t}\frac{\mathrm dt}{2\pi i} \\
& \times p(p\xi-\bar aHL)\mathrm e^{\pi L\xi}\mathrm e^{\pi i(t-s)/2}\left(\frac{2p}{H}\right)^{s+t} a_+^{-s}a_-^{-t} \\
& \times \frac{\Gamma(s+iL\xi)\Gamma(-s-\mu+1/2)\Gamma(-s+\mu+1/2)\Gamma(t-iL\xi)\Gamma(-t-\mu+1/2)\Gamma(-t+\mu+1/2)}
{\Gamma(1/2-iL\xi-\mu)\Gamma(1/2-iL\xi+\mu)\Gamma(1/2+iL\xi-\mu)\Gamma(1/2+iL\xi+\mu)}. 
\end{split}\end{equation}
We are now ready to perform the $p$-integral with a tiny shift in $p$ axis ($p \to p + i\varepsilon$). 
The residue theorem and perturbative ordering by the point separation $\epsilon$ give an analytic expression for the expectation value, 
\begin{equation}\label{Jzdiv}
\begin{split}
\braket{\hat J_z(x;\epsilon)} &= \frac{eaH^3}{4\pi^2}\biggl[-\frac{L}{3}(\log\epsilon+\log H+\frac{3}{2}+\gamma_E) - \frac{2}{15}L^3 \\
 & + \frac{\mu }{12\pi^3 L\sin(2\pi\mu)} \Bigl\{(45+4\pi^2(-2+3L^2+2\mu^2))\cosh(2\pi L) \\
 & \qquad\qquad\qquad\qquad\qquad - (45+8\pi^2(-1+9L^2+\mu^2))\frac{\sinh(2\pi L)}{2\pi L} \Bigl\}  \\
 & + \Re \Bigl[ \frac{iL}{16\sin(2\pi\mu)}\int_{-1}^{1}\mathrm d\xi (1-4\mu^2+(-7-12L^2+12\mu^2)\xi^2+20L^2\xi^4) \\
 & \times\left\{ 
 (\mathrm e^{2\pi L\xi}+\mathrm e^{-2i\pi\mu})\psi\left(\frac{1}{2}-iL\xi+\mu\right)-(\mathrm e^{2\pi L\xi}+\mathrm e^{2i\pi\mu})\psi\left(\frac{1}{2}-iL\xi-\mu\right) \right\} \Big] 
 + \mathcal O(\epsilon^1) \biggl], 
\end{split}
\end{equation}
where $\psi(z)=\Gamma^\prime(z)/\Gamma(z)$ denotes the digamma function and $\gamma_E$ is the Euler-Mascheroni constant. 
The covariant separation is expressed as $\epsilon^2 \equiv \epsilon^\mu\epsilon_\mu = a^2(-\Delta\eta^2+|\Delta\bm x|^2)/4$. 
Note that the separation in scale factor ($a_\pm \neq a$) must be preserved during calculation, otherwise it would lead to a wrong result. 
We have only a logarithmic divergence of the separation to be absorbed by the renormalization of the charge and the gauge field. 
In fact, we also have direction dependent divergent terms such as $(-\Delta\eta^2+|\Delta\bm x|^2)^{-2}\Delta z$. 
We can, however, eliminate them by adopting a rule that the limit $\Delta z\to 0$ must be taken in advance whenever the coincidence limit 
$\epsilon \to 0$ is taken. 

\subsection{Renormalization} 
The vacuum expectation value of the current operator must be placed at the right-hand side of 
the semiclassical Maxwell equation $F^{\mu\nu}_{\;\;\; ;\nu}(x) = - \braket{\hat J^\mu(x)}$. 

Renormalization prescription is required to deal with divergence. In our set up, only the $z$-component is relevant.  
We can use usual ansatz for renormalized field $A_{\mathrm R \, z}$ and charge $e_{\mathrm R}$ 
involving a divergent coefficient $C$ such as 
\begin{equation}
A_{\mathrm R \, z} = C A_{z}, \quad e_{\mathrm R} = C^{-1} e, 
\end{equation}
or instead of renormalizing $A_z$ we can introduce renormalized electric field strength $E_{\mathrm R} = CE$. 
Note that combination of the charge and the field is unchanged $e_{\mathrm R}E_{\mathrm R} = eE$, 
and $\braket{J_z(e,E) }= C \braket{J_z(e_{\mathrm R},E_{\mathrm R})}$. 
The minimal choice for $C$ is found to be 
\begin{equation}
C^2 = 1 + \frac{e^2}{24\pi^2}\log\epsilon.
\end{equation} 
This choice gets rid of only the $\log\epsilon$ term from (\ref{Jzdiv}). 
We can subtract the terms proportional to $L$ from large parenthesis $[\cdots]$ in (\ref{Jzdiv}) in addition to it. 
So the form of $C$ is given by 
\begin{equation}\label{renC}
C^2 = 1 + \frac{e^2}{24\pi^2}\left\{ \log\epsilon + (\mathrm{finite\;terms}) \right\}. 
\end{equation}

Some Physical condition is required to determine the finite part in ($\ref{renC}$). 
We adopt the requirement that the renormalized current must vanish in massive scalar limit ($m^2\gg E,\,H^2$), 
\begin{equation}\label{rencondi}
\lim_{M\to\infty}\braket{\hat J_z}_{\mathrm{ren}} = 0. 
\end{equation}
The asymptotic behavior of the digamma function $\psi(z)\sim \log(z)-1/(2z)+\mathcal O(z^{-2})$ is useful to find nonvanishing terms in massive limit 
\begin{equation}\label{Jdiv}
\braket{\hat J_z} \xrightarrow{M\gg1,\,L\ll1} -\lim_{\epsilon\to 0} \frac{eaH^3}{4\pi^2}\frac{L}{3}
\left(\log\epsilon + \log m + \gamma_E + 3/2 \right). 
\end{equation}
This tells us the minimal form of the finite terms in ($\ref{renC}$) and we can obtain the renormalized current 
\begin{equation}\label{Jzren}
\begin{split}
&\braket{\hat J_z(x)} \\
 & = \frac{eaH^3}{4\pi^2}\biggl[ \frac{L}{3}\log M - \frac{2}{15}L^3 
 + \frac{\mu }{12\pi^3 L\sin(2\pi\mu)} \Bigl\{(45+4\pi^2(-2+3L^2+2\mu^2))\cosh(2\pi L) \\
 & \qquad\qquad\qquad\qquad\qquad - (45+8\pi^2(-1+9L^2+\mu^2))\frac{\sinh(2\pi L)}{2\pi L} \Bigl\}  \\
 & + \Re \Bigl[ \frac{iL}{16\sin(2\pi\mu)}\int_{-1}^{1}\mathrm d\xi (1-4\mu^2+(-7-12L^2+12\mu^2)\xi^2+20L^2\xi^4) \\
 & \qquad\qquad \times\left\{ 
 (\mathrm e^{2\pi L\xi}+\mathrm e^{-2i\pi\mu})\psi\left(\frac{1}{2}-iL\xi+\mu\right)-(\mathrm e^{2\pi L\xi}+\mathrm e^{2i\pi\mu})\psi\left(\frac{1}{2}-iL\xi-\mu\right) \right\} \Big] \biggl]. 
\end{split}
\end{equation}

Thus we have reached the same expression as obtained by Kobayashi and Afshordi 
\cite{kobayashi2014schwinger} using the adiabatic regularization up to the second order.

\section{Properties of the result}\label{Sec:result}
\begin{figure}[htb]
 \begin{center}
  \includegraphics[width=120mm]{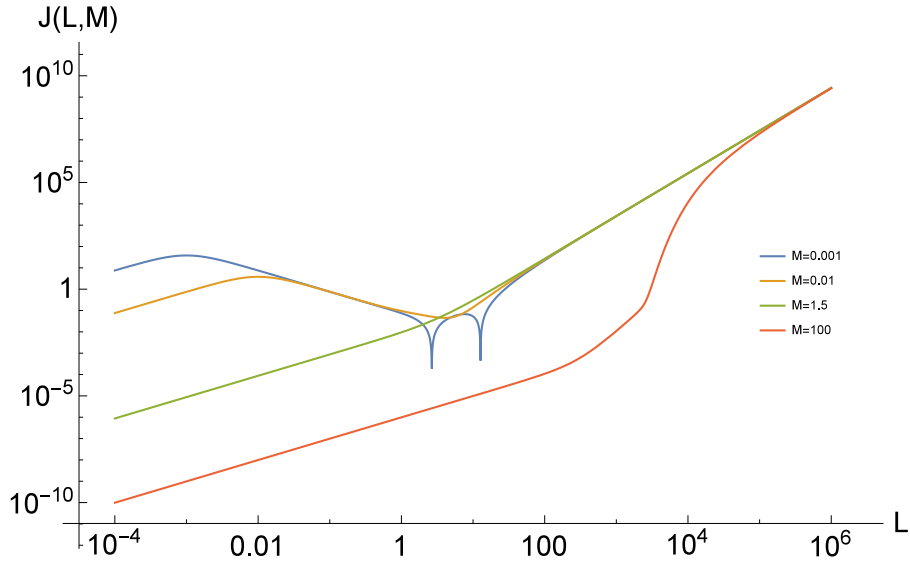}
 \end{center}
 \caption{Absolute value of the renormalized current $J = \braket{J}/eaH^3$ is shown as a function of $L=eE/H^2$. 
 Each line corresponds to different mass parameter $M=m/H$. 
 Negative current is observed in $L=1\sim10$ for $M=0.001$ case. }
 \label{fig:current}
\end{figure}

It is remarkable that our result agrees with the previous one 
\cite{kobayashi2014schwinger}, and worthwhile to list its physical
significance for self-containedness.
Note that the dimensionless current defined as 
\begin{equation}
J = J(L,M)\equiv \dfrac{\braket{\hat J_z}}{eaH^3}, 
\end{equation}
is a function of $L$ and $M$. 
The graph of $J(L,M)$ as a function of the electric field strength $L$ is shown in Fig.~\ref{fig:current} for different values of the mass parameter $M$.

\subsection{Limiting behaviors}

First let us consider various limiting behaviors.

\subsubsection{Weak electric field regime}$eE\ll m^2,H^2$

\begin{figure}[htb]
 \begin{center}
  \includegraphics[width=110mm]{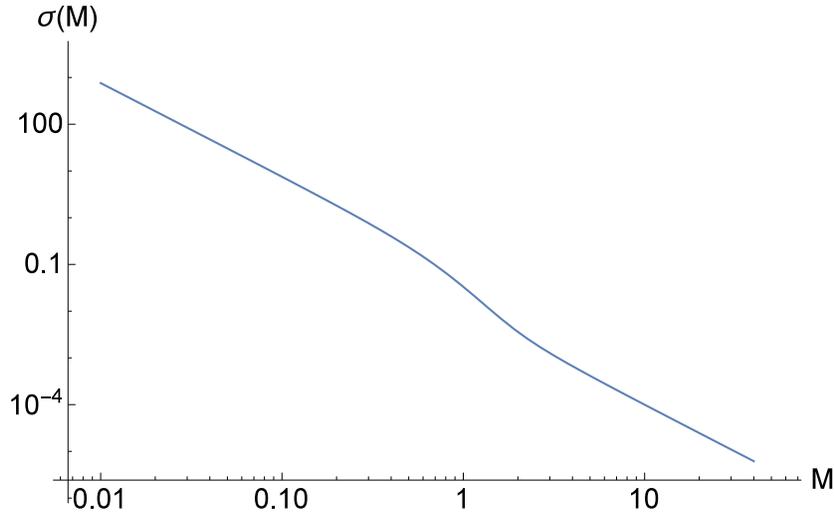}
 \end{center}
 \caption{Conductivity of dS spacetime}
 \label{fig:conductivity}
\end{figure}

The renormalized current in this regime is expressed as 
\begin{equation}\label{Jmassivelimit}
J \xrightarrow{L\to0} \frac{L}{12\pi^2} 
\left\{ \log M +\frac{8\pi}{3}\frac{\mu_0(\mu_0^2-1)}{\sin(2\pi\mu_0)}-\frac{1}{2}\left[\psi\left(\frac{1}{2}+\mu_0\right)+\psi\left(\frac{1}{2}-\mu_0\right)\right]\right\}, 
\end{equation}
where $\mu_0=\sqrt{9/4-M^2}$. 
The dimensionless conductivity $\sigma(M) = J/L\vert_{L\to0}$ is plotted in Fig.~\ref{fig:conductivity}. 
The scaling of $\sigma(M)$ is given by 
\begin{equation}\label{conductivity}
\sigma(M)\rightarrow \left\{
\renewcommand{\arraystretch}{1.85}
\begin{array}{cc}
 \dfrac{3}{4\pi^2 M^2}  & (M\ll1) \\
 \left(\dfrac{7}{72\pi^2M^2} + \mathcal O(M^{-4})\right) - \mathrm e^{-2\pi M} \left(\dfrac{4}{9\pi}M^3 + \mathcal O(M^1)\right)& (M\gg1)
\end{array}
\renewcommand{\arraystretch}{1}
\right. . 
\end{equation}
There is a strong $M^{-2}$ enhancement for the small scalar mass. 
This is a four dimensional analog of the two dimensional IR hyperconductivity reported in \cite{1475-7516-2014-04-009}. 
We can also see the $M^{-2}$ scaling for the massive scalar but the coefficient is slightly changed. 

The exponentially suppressed term ($\propto \mathrm e^{-2\pi M}$) in the large mass regime must exist naturally 
because the standard Bogoliubov calculation gives the number density of the scalar particles in dS spacetime 
$n\sim H^4 (\mathrm e^{2\pi M}-1)^{-1}$ which means the exponential suppression of heavy particles. 
Nevertheless, we also have inexplicable terms which are not protected by exponential factor in the conductivity.

Of course, there is room for changing the renormalization fixing without breaking the condition $\braket{J_z}_{\mathrm{ren}} \to 0$ for $M\to\infty$. 
If one naively tried to remove the $M^{-2}$ term in (\ref{conductivity}), it would cause a huge IR correction 
and even worse negativity to the renormalized current. 
Furthermore, if all the unprotected terms should be subtracted from the current $J$, the second term in (\ref{Jmassivelimit}) 
would be left results in a discontinuity at $M=\sqrt 2$. 
\footnote{
This mass parameter corresponds to the conformal coupling $\xi R\phi^2$ in dS spacetime, $\xi=1/6$ and $R=12H^2$. 
Thus this is conformally equivalent to a massless scalar field in Minkowski spacetime. 
Thus this discontinuity (or divergence) at $M=\sqrt 2$ might be physically reasonable. 
} 
It is also obvious that $\sigma$ would be negative for $M\gg1$ in such a treatment. 
For these reasons, we do not consider changing the renormalization condition. 

Note also that this $7/(72\pi^2 M^{2})$ term corresponds to the fourth order adiabatic term. 
The terms in (\ref{Jdiv}) correspond to the zeroth and second order adiabatic subtraction terms. 
We can expect that the formal infinite order adiabatic subtraction of the terms proportional to $L$ (WKB is an asymptotic expansion) 
would result in the removal of the exponentially unprotected behavior in massive limit.

\subsubsection{Strong electric field or weak curvature regime}$eE\gg m^2,\,H^2$, or $H^2\ll m^2,\, eE$

In this limit, the $L^3$ term in the first line of (\ref{Jzren}) and the integration of the digamma functions cancels each other. 
We find
\begin{equation}
J \simeq \frac{L^2}{12\pi^3} \mathrm{sgn}(L) \mathrm e^{-\frac{\pi M^2}{|L|}}, 
\end{equation}
and recover the Schwinger's famous suppression factor $\exp(-\pi m^2/eE)$. 
There is no mass dependence for $L\to\infty$. All the lines in Fig.~\ref{fig:current} converge at infinity. 
Equivalently, this is expressed in terms of dimensionful quantities in Minkowski limit as
\begin{equation}
\braket{J_z}_{\mathrm{ren}} \xrightarrow{H\to0} \frac{e}{12\pi^2}(eE)^2 \frac{1}{H} e^{-\frac{\pi m^2}{eE}}, 
\end{equation}
where asymptotic analysis reveals that no $\mathcal O(H^0)$ term appears. The divergence $H^{-1}$ is due to the lack of the cosmic dilution in the Minkowski spacetime. 
The particles produced at $t=-\infty$ contributes to the current expectation value forever, so $H^{-1}$ must be replaced by 
some regulator such as $(t-t_0)$ with $t_0$ being the turn-on time of the electric field. 
This prescription is justified because the differentiation $\frac{\mathrm d}{\mathrm d\eta} \braket{J}$ is finite when $H\to0$. 
Note that the conformal time $\eta$ is identical to the cosmic time $t$ in this limit as we are taking the scale factor $a=(1-H\eta)^{-1}$. 
This behavior corresponds to the result obtained in Minkowski space. 
This linear growth of the current in time was shown in \cite{Anderson:2013ila}.

\subsection{Negativity of the current}
\begin{figure}[htb]
 \begin{center}
  \includegraphics[width=110mm]{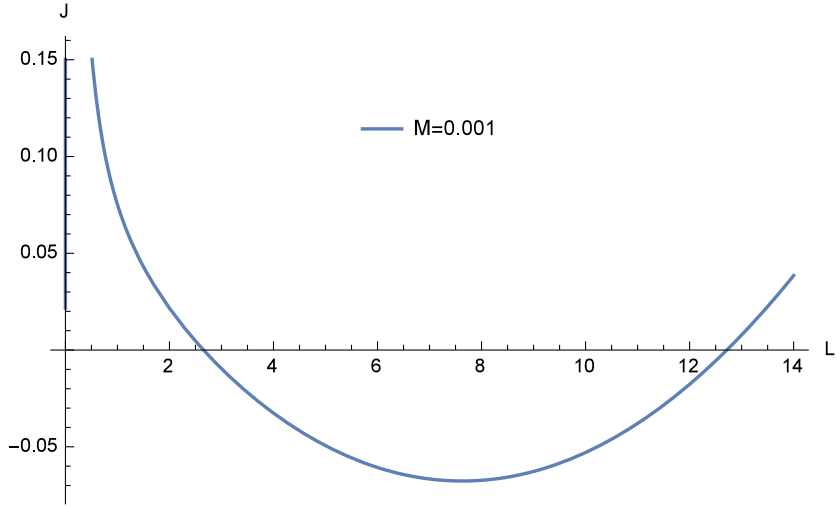}
 \end{center}
 \caption{Example of the negativity of the current $J(L,M)$. 
 IR enhancement around $L\sim M$ is not shown here (see the logarithmic plot in Fig.~\ref{fig:current}). 
 There are three zeros of $J$, two of them ($L=0$ and $L\sim 13$) are stable, the other ($L\sim2.5$) is unstable. 
 }
 \label{fig:zeroofJ}
\end{figure}

Noteworthy is the existence of the negative current $J<0$ around $L\sim \mathcal O(1)$ for small mass regime $M\lesssim 10^{-3}$. 
Typical situation is depicted in Fig.~\ref{fig:zeroofJ}. 
$L=0$ is a trivial zero of $J(L,M)$. Two more zeros of the current appear in $L>0$ and $J$ is negative between them. 

The positive current causes negative backreaction to the background electric field as expected. 
The negative current conversely enhances the background electric field. 
The current-electric field system can be seen as a sort of feedback system and stability analysis is easy. 
The first (nontrivial) zero of the current corresponds to the so-called diverging point. 
No backreaction occurs at this point, however, small deviation from this point induces positive feedback which enhances the deviation. 
The second zero (and also the trivial zero) is a stable point of the system. Small deviations are pulled back to this point. 

Note that negativity of the induced current does not mean fatal instability of the system, as it occurs only in a finite range of the electric field 
and the induced current recovers its positivity as $L$ increases. The position of the second zero is numerically given by 
$L = -\frac{1}{3} \log M$ for $M\ll1$ indeed.

\begin{figure}[htb]
 \begin{center}
  \includegraphics[width=110mm]{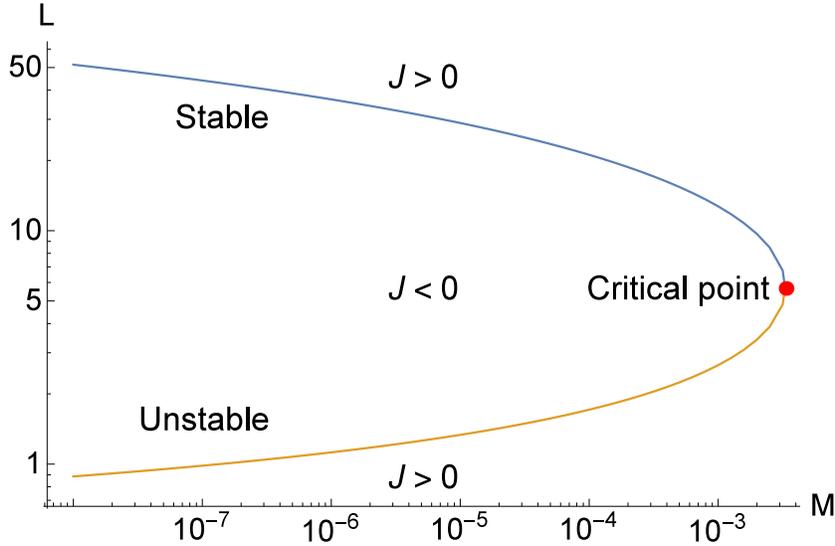}
 \end{center}
 \caption{Linear stability analysis of the current-electric field system. 
 The upper (blue) line represents the trajectory of stable points and the lower (orange) line represents that of diverging points. 
 The red dot represents the critical point $(L_c,M_c) = (5.7,0.0033)$. 
 }
 \label{fig:stability}
\end{figure}

In Fig.~\ref{fig:stability}, we show the position of the zeros of $J(L,M)$ in $L$-$M$ plane, which
also serves as a phase diagram of the current versus electric field.
Each line represents the zeros of the current. The lower (upper) line corresponds to the first (second) zero of the current $J(L,M)$. 
The negativity happens in the region between the two lines. 
There exists a critical mass $M_c \sim 0.0033$, the maximum value of the scalar mass which can cause the negativity of the current. 
We also emphasize that the negativity is usually mild compared to the IR hyperconductivity which occurs coincidently 
and the $J\propto L^2$ behavior in the strong field regime as we mentioned in previous subsection. 

Although we have convinced ourselves that the negative induced 
current is not dangerous as we intuitively thought, its interpretation is another problem. 
We must remember that the particle description is correct only in the semiclassical regime, say, $L^2+M^2\gg1$. 
Naively taken, small mass but large electric field regime can satisfy this condition and 
should be described well by the semiclassical approximation, but it turns out not to be the case. 
The appearance of the exponentially unprotected terms in weak electric field but massive limit (\ref{conductivity}) 
is the counterpart of the breakdown of the semiclassical description.

\section{Discussion}\label{Sec:discussion}
In this section, we try to clarify the intuitive physical interpretation of the results summarized in the previous section. 

We begin with the semiclassical description for the induced current $\braket{J_z}$. In the particle picture, $\braket{J_z}$ can be 
split into two parts (see also \cite{1475-7516-2014-04-009}) as 
\begin{equation}
\braket{J_z} = J_{\mathrm{pairs}} + J_{\mathrm{vac}}, 
\end{equation}
where $J_{\mathrm{pairs}}$ is a contribution due to the kinetic motion of the semiclassical particles 
and given by $J_{\mathrm{pairs}}=2e \int v\mathrm dn$ ($v$ is the velocity and 
$\mathrm dn$ is the differential number density of carriers). 
$J_{\mathrm{vac}}$ is the vacuum current which flows to satisfy the local current conservation low 
when the particle pair is produced out of the vacuum. The vacuum current connects the pair particles. 
In Minkowski spacetime, the positiveness of the semiclassical current is obvious. 
However, it is not so trivial in de Sitter spacetime due to the dynamics of the charged particles and the spacetime topology. 
When the charged particles move along the electric field, $J_{\mathrm{pairs}}$ is always positive, but they can move against the 
electric field (in upstream direction) in de Sitter spacetime. This is due to the effect of the rapid cosmic expansion. 
Of course, $J_{\mathrm{pairs}}$ itself is expected to be positive because the particle production in the usual (down-stream) direction 
is more likely than that in the opposite (up-stream) direction even in de Sitter spacetime. 
Another difference from Minkowski appears in $J_{\mathrm{vac}}$. 
Since four-dimensional de Sitter spacetime has $\mathbb{R} \times S^3$ topology, there is a detour 
which links two spatially separated points. Thus, the vacuum current can make a detour and contribute negatively, 
while the vacuum current in a shortcut way is always positive. 
Since a scalar field with a small mass has a long Compton wavelength, it can detect the global structure of spacetime. 
If this explanation for the negative current is pertinent, the negativity of the induced current might be found even 
in a lower dimensional setup such as quantum field theory on a circle. 

When it comes to the terms without the exponential mass suppression in \eqref{conductivity}, we cannot make any satisfactory interpretations 
so far as they are contributions beyond those obtained by the semiclassical approximation or the instanton analysis \cite{Brown:2015kgj}.

\section{Conclusion}\label{Sec:conclusion}
In the present paper, we have calculated the vacuum expectation value of a charged scalar current in the presence of a constant homogeneous 
electric field along $z$-direction in de Sitter spacetime using the point-splitting regularization scheme in a covariant and 
gauge-invariant manner. This enabled us to do renormalization explicitly in the equation of motion. 
The only divergence we have encountered is the logarithmic divergence which can be absorbed into the kinetic term of the gauge field in a conventional fashion. 
In a previous calculation done with the momentum cutoff technique \cite{kobayashi2014schwinger}, 
there was also a quadratic divergence which was an obstacle to the gauge-invariant renormalization. 
In \cite{kobayashi2014schwinger}, the adiabatic subtraction, in which WKB expansion was adopted to imitate 
the large momentum behavior of the mode function, was applied. 
Here we have imposed the renormalization condition (\ref{rencondi}) instead of employing the adiabatic subtraction. 
Interestingly, the result of the minimal subtraction eliminating only the terms in (\ref{Jdiv}) and 
that with the adiabatic subtraction show the perfect agreement. 
This remarkable fact strongly suggests the correctness of the result. 

We have also investigated the properties and the consequences of the renormalized current (\ref{Jzren}). 
We have found two kinds of the breakdown of the semiclassical approximation. 
One is the term which is \emph{not} suppressed by the exponential factor $\mathrm e^{-2\pi m/H}$ 
and it appears in the massive and weak electric field limit, $eE/H^2\ll1,\,m/H\gg1$. 
The conventional Bogoliubov calculation indicates that all the terms in this limit should be protected by this exponential factor. 
The limiting behavior of the (dimensionless) conductivity $\sigma = \braket{J}_{\mathrm{ren}}/(e^2aHE)|_{E\to0}$ in (\ref{conductivity}), 
however, contains the unprotected terms. 

The other is the negativity of the renormalized current which shows up in tiny mass regime $m\lesssim0.003 H$ 
that has already been discovered by Kobayashi and Afshordi \cite{kobayashi2014schwinger} using the adiabatic regularization scheme. 
It is natural that one might think the negative current of an artifact of the renormalization scheme. 
Indeed, the adiabatic subtraction scheme breaks down in the IR limit 
because WKB (adiabatic) approximation is not correct in this regime even though it removes the UV divergences. 
We have, however, found that the outcome has nothing to do with the accuracy of the WKB approximation, 
so that we can say that such a criticism does not apply. 
Therefore, we have to take these strange phenomena seriously. 

It should be noted that the expression for the renormalized current \eqref{Jzren} apparently has a divergence in the massless limit.
This is not the problem of our analysis, but merely an outcome of the fact that the vacuum state for an exactly massless charged field 
in an electromagnetic background is unstable. 
Of course, there is no divergence in the large mass limit by construction. 

We have argued that the uncertainty of the renormalization in curved spacetime comes from the lack of knowledge of 
the correct behaviors of the quantum fields in some asymptotic region. 
The only reliable behavior is the asymptotics in Minkowski limit, but it does not fully fix the renormalization condition, 
although the semiclassical properties are reproduced in this limit, namely, for the cases $L\gg 1$ or $H\ll 1$. 
Again, this is not the problem of our analysis but rather originates in the lack of information in this curved spacetime. 
Conversely, the choice of the renormalization condition does not affect the behavior in the flat spacetime. 

\acknowledgments
This work was supported by JSPS KAKENHI, 
Grant-in-Aid for JSPS Fellows 15J09390 (TH), 
Grant-in-Aid for Scientific Research 15H02082 (JY), 
Grant-in-Aid for Scientific Research on Innovative Areas 15H05888 (JY).

\bibliographystyle{JHEP.bst}
\bibliography{ref}

\providecommand{\href}[2]{#2}\begingroup\raggedright\begin{thebibliography}{10}

\bibitem{birrell1984quantum}
N.~Birrell and P.~Davies, {\em Quantum {F}ields in {C}urved {S}pace}.
\newblock Cambridge Monographs on Mathematical Physics. Cambridge University
  Press, 1984.

\bibitem{mukhanov2007introduction}
V.~Mukhanov and S.~Winitzki, {\em Introduction to Quantum Effects in Gravity}.
\newblock Cambridge University Press, 2007.

\bibitem{doi:10.1142/S0218271815300256}
K.~Sato and J.~Yokoyama, {\it Inflationary cosmology: First 30+ years},  {\em
  International Journal of Modern Physics D} {\bf 24} (2015), no.~11 1530025.

\bibitem{plaga1995detecting}
R.~Plaga, {\it Detecting intergalactic magnetic fields using time delays in
  pulses of $\gamma$-rays},  1995.

\bibitem{0004-637X-682-1-127}
K.~Ichiki, S.~Inoue, and K.~Takahashi, {\it Probing the nature of the weakest
  intergalactic magnetic fields with the high-energy emission of gamma-ray
  bursts},  {\em The Astrophysical Journal} {\bf 682} (2008), no.~1 127.

\bibitem{neronov2010evidence}
A.~Neronov and I.~Vovk, {\it Evidence for strong extragalactic magnetic fields
  from {F}ermi observations of {T}ev blazars},  {\em Science} {\bf 328} (2010),
  no.~5974 73--75.

\bibitem{2041-8205-747-1-L14}
I.~Vovk, A.~M. Taylor, D.~Semikoz, and A.~Neronov, {\it Fermi/{LAT}
  {O}bservations of 1{ES} 0229+200: {I}mplications for {E}xtragalactic
  {M}agnetic {F}ields and {B}ackground {L}ight},  {\em The Astrophysical
  Journal Letters} {\bf 747} (2012), no.~1 L14.

\bibitem{2041-8205-744-1-L7}
K.~Takahashi, M.~Mori, K.~Ichiki, and S.~Inoue, {\it Lower {B}ounds on
  {I}ntergalactic {M}agnetic {F}ields from {S}imultaneously {O}bserved
  {G}e{V}-{T}e{V} {L}ight {C}urves of the {B}lazar {M}rk 501},  {\em The
  Astrophysical Journal Letters} {\bf 744} (2012), no.~1 L7.

\bibitem{tashiro2014search}
H.~Tashiro, W.~Chen, F.~Ferrer, and T.~Vachaspati, {\it Search for {CP}
  violating signature of intergalactic magnetic helicity in the gamma-ray sky},
   {\em Monthly Notices of the Royal Astronomical Society: Letters} {\bf 445}
  (2014), no.~1 L41--L45.

\bibitem{Akahori01062014}
T.~Akahori, K.~Kumazaki, K.~Takahashi, and D.~Ryu, {\it Exploring the
  intergalactic magnetic field by means of faraday tomography},  {\em
  Publications of the Astronomical Society of Japan} {\bf 66} (2014), no.~3
  [\href{http://arxiv.org/abs/http://pasj.oxfordjournals.org/content/66/3/65.full.pdf+html}{{\tt
  http://pasj.oxfordjournals.org/content/66/3/65.full.pdf+html}}].

\bibitem{Schnitzeler01112010}
D.~H. F.~M. Schnitzeler, {\it The latitude dependence of the rotation measures
  of nvss sources},  {\em Monthly Notices of the Royal Astronomical Society:
  Letters} {\bf 409} (2010), no.~1 L99--L103,
  [\href{http://arxiv.org/abs/http://mnrasl.oxfordjournals.org/content/409/1/L99.full.pdf+html}{{\tt
  http://mnrasl.oxfordjournals.org/content/409/1/L99.full.pdf+html}}].

\bibitem{ratra1992cosmological}
B.~Ratra, {\it Cosmological 'seed' magnetic field from inflation},  {\em The
  Astrophysical Journal} {\bf 391} (1992) L1--L4.

\bibitem{PhysRevD.37.2743}
M.~Turner and L.~Widrow, {\it Inflation-produced, large-scale magnetic fields},
   {\em Phys. Rev. D} {\bf 37} (May, 1988) 2743--2754.

\bibitem{PhysRevD.69.043507}
K.~Bamba and J.~Yokoyama, {\it Large-scale magnetic fields from inflation in
  dilaton electromagnetism},  {\em Phys. Rev. D} {\bf 69} (Feb, 2004) 043507.

\bibitem{PhysRevD.70.083508}
K.~Bamba and J.~Yokoyama, {\it Large-scale magnetic fields from dilaton
  inflation in noncommutative spacetime},  {\em Phys. Rev. D} {\bf 70} (Oct,
  2004) 083508.

\bibitem{martin2008generation}
J.~Martin and J.~Yokoyama, {\it Generation of large scale magnetic fields in
  single-field inflation},  {\em Journal of Cosmology and Astroparticle
  Physics} {\bf 2008} (2008), no.~01 025.

\bibitem{demozzi2009magnetic}
V.~Demozzi, V.~Mukhanov, and H.~Rubinstein, {\it Magnetic fields from
  inflation?},  {\em Journal of Cosmology and Astroparticle Physics} {\bf 2009}
  (2009), no.~08 025.

\bibitem{schwinger1951gauge}
J.~Schwinger, {\it On gauge invariance and vacuum polarization},  {\em Physical
  Review} {\bf 82} (1951), no.~5 664.

\bibitem{PhysRevD.49.6343}
J.~Garriga, {\it Pair production by an electric field in (1+1)-dimensional de
  {S}itter space},  {\em Phys. Rev. D} {\bf 49} (Jun, 1994) 6343--6346.

\bibitem{1475-7516-2014-04-009}
M.~B. Fr{\"o}b, J.~Garriga, S.~Kanno, M.~Sasaki, J.~Soda, T.~Tanaka, and
  A.~Vilenkin, {\it Schwinger effect in de {S}itter space},  {\em Journal of
  Cosmology and Astroparticle Physics} {\bf 2014} (2014), no.~04 009.

\bibitem{Cai:2014qba}
R.-G. Cai and S.~P. Kim, {\it {One-Loop Effective Action and Schwinger Effect
  in (Anti-) de Sitter Space}},  {\em JHEP} {\bf 09} (2014) 072,
  [\href{http://arxiv.org/abs/1407.4569}{{\tt arXiv:1407.4569}}].

\bibitem{kobayashi2014schwinger}
T.~Kobayashi and N.~Afshordi, {\it Schwinger effect in 4{D} de {S}itter space
  and constraints on magnetogenesis in the early universe},  {\em Journal of
  High Energy Physics} {\bf 2014} (2014), no.~10 1--36.

\bibitem{Stahl:2015gaa}
C.~Stahl, E.~Strobel, and S.-S. Xue, {\it {Fermionic current and Schwinger
  effect in de Sitter spacetime}},  \href{http://arxiv.org/abs/1507.0168}{{\tt
  arXiv:1507.0168}}.

\bibitem{Bavarsad:2016cxh}
E.~Bavarsad, C.~Stahl, and S.-S. Xue, {\it {Scalar current of created pairs by
  Schwinger mechanism in de Sitter spacetime}},
  \href{http://arxiv.org/abs/1602.0655}{{\tt arXiv:1602.0655}}.

\bibitem{Hayashinaka:2016qqn}
T.~Hayashinaka, T.~Fujita, and J.~Yokoyama, {\it {Fermionic Schwinger effect
  and induced current in de Sitter space}},
  \href{http://arxiv.org/abs/1603.0416}{{\tt arXiv:1603.0416}}.

\bibitem{dewitt1975quantum}
B.~S. DeWitt, {\it Quantum field theory in curved spacetime},  {\em Physics
  Reports} {\bf 19} (1975), no.~6 295--357.

\bibitem{gradshteyn2007}
I.~S. Gradshteyn and I.~M. Ryzhik, {\em Table of integrals, series, and
  products}.
\newblock Elsevier/Academic Press, Amsterdam, seventh~ed., 2007.

\bibitem{Anderson:2013ila}
P.~R. Anderson and E.~Mottola, {\it {Instability of global de Sitter space to
  particle creation}},  {\em Phys. Rev.} {\bf D89} (2014) 104038,
  [\href{http://arxiv.org/abs/1310.0030}{{\tt arXiv:1310.0030}}].

\bibitem{Brown:2015kgj}
A.~R. Brown, {\it {Schwinger pair production at nonzero temperatures or in
  compact directions}},  \href{http://arxiv.org/abs/1512.0571}{{\tt
  arXiv:1512.0571}}.

\end{thebibliography}\endgroup

\end{document}